# Beam Collimation Using an Anisotropic Metamaterial Slab without Any Nanometer-sized Aperture

Shou Zhang,[1] Guohui Li,[1] Yanxia Cui,[1,2,*] Feng Zhang,[3] Sailing He,[3] Yuying Hao,[1,*] and Furong Zhu[1,2,*]

[1]*Key Lab of Advanced Transducers and Intelligent Control System of Ministry of Education, College of Physics and Optoelectronics, Taiyuan University of Technology, Taiyuan 030024, China*

[2]*Department of Physics and Institute of Advanced Materials, Hong Kong Baptist University, Kowloon Tong, Hong Kong*

[3]*Centre for Optical and Electromagnetic Research, Zhejiang University, Hangzhou, Zhejiang 310058, China*

[*]Corresponding authors: *yanxiacui@gmail.com*, *haoyuying@tyut.edu.cn*, *frzhu@hkbu.edu.hk*

## Abstract

Plasmonic beam collimation effect has been thoroughly investigated based on the well-known nanometer-scale bull's eye structure formed by complex and high-cost fabrication processes. In this work, we report our effort for attaining beam collimation using an anisotropic metamaterial (AMM) slab that consists of a stack of alternating metal/dielectric layers and an integrated top metal grating. The results show that AMM slab allows creating the beam collimation effect similar to that of the bull's eye structure, an enabling technology for practical application due to its simple architecture and cost benefits. The excitation of surface plasmons at the AMM/air interface is derived. The structure of the AMM slab and its impact on beaming performance were analyzed using the effective medium theory and Finite Element Method.

Beam shaping is of great importance in laser technology, light emission devices and optical imaging process [1, 2]. Conventionally, it is realized by sets of bulky optical lens or aspherical lens, facing challenges to integrate the optical components into the host system. In the last decade, new device concepts and ideas for beam shaping are emerging due to the rapid development of surface plasmon (SP) theory [3, 4] and nanofabrication techniques [5, 6]. For example, plasmonic alternatives with SP wave excited on metallic nanostructures can manipulate light in subwavelength scale, enabling the miniaturization of conventional devices. SP has attracted intensive attention and has already been applied in many areas, such as beam shaping [7-23], signal sensing [24, 25], design of absorbers [26, 27], waveguide optics [28, 29], photovoltaic and light emission devices [30-32], etc.

The so-called bull's eye structure [7], composed of a metallic nano-aperture with surrounding periodical shallow engraved grooves, is the pioneering SP-based architecture applied for beam shaping. The whole device occupies no more than one cubic micrometer, much more compact in comparison with a traditional lens system. The optimized beaming effect along the on-axis direction requires that the excited SPs from the slit and the textured grooves in the exit plane must be approximately in phase [8]. Beaming will perform along off-axis directions when asymmetric factors are applied during the excitation of SPs [10, 19, 21, 23]. Later, this miniaturized beaming system was introduced into semiconductor lasers aiming for reducing the divergence angle of emission [11, 14, 16]. By proper design, different beam shaping effects, *e.g.*, focusing [9, 12, 13, 20], can be realized. Such phenomenon has already been demonstrated for waves at THz [16] and microwave frequency ranges [17]. However, in most of these designs, an opaque metallic film with milled apertures is a required, for isolating the output space from the incident space so that the output radiation is only determined by the modes excited in the exit plane. The apertures allow light passing from the incident to the output space. However, the dimensions of the apertures should be in the scale of nanometers for visible light frequencies, bringing many difficulties in fabrication. The focused ion beam method is used to make the nanometer-scale bull's eye structures [7, 11, 16, 19, 22], which is an expensive and time-consuming fabrication process.

In this report, we present the results of a plasmonic beam collimator that consists of a group of periodical shallow grooves and a stack of alternating planar metal/dielectric layers. The collimator possesses a very similar angular divergence characteristic for visible light compared to that of the bull's eye structure. In this design, the metal layer becomes transmissive after the insertion of multiple dielectric layers, yielding the beam shaping without the nanometer-sized passway. The metal/dielectric composite layer, as a whole, works as a substrate in terms of anisotropic metamaterials (AMMs) [33], which can be fabricated using the electron-beam evaporation method [34]. By this way, the processing difficulty of the plasmonic beam collimator can be reduced. As known, AMMs have already been applied in many areas, including superlenses [35], absorbers [26], beam splitters [36], etc.; however, the use of AMMs for application in beam collimation has not been reported yet. The propagating SP mode excited on such an AMM substrate, which distinctively differs from that excited on a flat metal surface, is derived in this work. The effects of the geometrical parameters of the AMM slab design on the beam collimation are analyzed.

**Results**

**AMM based beam collimation design.** The schematic diagrams of the two-dimensional (2D) and three-dimensional (3D) configurations of the proposed plasmonic beam collimator are shown in Fig. 1(a) and 1(b). The top silver (Ag) grating with the periodicity of $p$, groove height of $h$ and width of $w$ is placed on top of the AMM slab. The stack of metal/dielectric layers on a glass substrate, made with alternating Ag and molybdenum oxide ($MoO_3$) films, functioned as the AMM composite layer. $t_1$ and $t_2$ are the thicknesses of Ag and $MoO_3$ layers, and the total thickness of the AMM composite layer is denoted by $H_{AMM}$. Overall there are 7 alternating layers in our design and both the top and bottom layers of the AMM part are made of silver. For such a 2D plasmonic configuration, the SP mode is only excited at TM polarization with the magnetic field $H_z$ perpendicular to the cross-sectional plane (*i.e.*, the *x-y* plane). A Gaussian beam of waist radius ($w_b$) at wavelength ($\lambda_0$) at TM polarization incidents from the glass substrate onto the device. The wavelength is fixed to $\lambda_0$ = 632.8 nm, at which the complex relative permittivity of Ag is $\varepsilon_m$ = -15.9 + 1.07i [37] and that of MoO$_3$ is $\varepsilon_d$ = 3.74 + 0.028i obtained by experimental measurements. The default incident Gaussian beam has a waist radius of $w_b$ = 1.5 μm. The default structural parameters are $p$ = 554 nm, $w/p$ = 0.2, $h$ = 36 nm, $H_{AMM}$ = 70 nm, and $t_2/t_1$ = 0.9. It is noted that the total number of the pairs of grooves should not be too small [8]; here 20 pairs of grooves are considered.

**Demonstration of the AMM based beam collimation effect.** A thick metal plate tends to block light transmission, thus the bull's eye structure requires an aperture for guiding light from the incident space to the output space. But in our design, the metal layer becomes transmissive after the insertion of multiple dielectric layers. Fig. 2(a) shows the $H_z$ field distribution when the incident Gaussian beam with a waist radius of 1.5 μm passes through the proposed structure. First, it is seen that part of the incident light can successfully transmit through the designed structure. Besides, it is noticed that, by our proposed structure, the radius width of the output beam becomes much larger than that of the incident Gaussian beam, reflecting that the divergence property of the beam has been improved apparently. The comparison of the normalized angular dependent far-field intensities ($I_N \sim \theta$) obtained with and without the beam collimator is shown in Fig. 3(b). For better illustration, we also extracted the far-field distribution of $I_N \sim \theta$ for the referred bull's eye structure which was reported in Ref. [7] and redraw it in Fig. 3(a). In the inset of Fig. 3(a), the reference structure is also displayed. We define an indicator of the beam emission characteristic, the half width of emission angle, denoted by $\Delta\theta$, which relates to the angular range in which the field intensity reduces from the maximum to its half. For the reference structure, $\Delta\theta$ of the output beam is ~3 °. Here, by our design, $\Delta\theta$ of the output beam can reach the same level, ~3.36 °, reduced by a factor of 48.4% when compared with the original Gaussian beam ($\Delta\theta$ = 6.51 °). The side-lobes at larger angle in the far-field distribution curve of the AMM collimator are not identified in Fig. 3(a) because the applied incidence in Ref. [7] has a large waist width, as shown by the star-line in Fig. 4(a). It is clear that the AMM slab thus possesses the beam collimation effect similar to that of the well-known bull's eye structure.

It is straightforward to consider the parallel metal/dielectric composite layer with a feature size in deep subwavelength scale as a bulk AMM plate of which the diagonal elements of the relative permittivity are ($\varepsilon_{\parallel}$, $\varepsilon_{\perp}$, $\varepsilon_{\parallel}$). $\varepsilon_{\parallel}$ represents the equivalent relative permittivity along the metal/dielectric interface, here, the x-z plane, while $\varepsilon_{\perp}$ corresponds to that perpendicular to the metal/dielectric interface, here, along y axis. According to the effective medium theory, the equivalent parameters can be written as $\varepsilon_{\parallel} = f\varepsilon_m + (1-f)\varepsilon_d$, and $\varepsilon_{\perp} = 1/[f/\varepsilon_m + (1-f)/\varepsilon_d]$, where f is the filling fraction of the metal layer [i.e., $t_1/(t_1+t_2)$] [38]. For our case, $\varepsilon_{\parallel}$ = -6.60 + 0.58i, and $\varepsilon_{\perp}$ = 10.7 + 0.36i. In the following simulation, we regard the bottom part of our proposed beam collimator as a bulk AMM medium with equivalent material parameters. The corresponding $H_z$ field distribution and the normalized angular dependent far-field distributions of the equivalent structure are shown in Fig. 2(b) and Fig. 3(c), respectively. It is observed that there are negligible differences between the field distributions demonstrated by our real structure and the equivalent structure. The far-field emission profile of the output beam for the equivalent structure also indicates that the beam collimation effect is realized by the composition of such a bulk effective medium substrate and the on-top shallow periodical metallic grating. Quantitatively, $\Delta\theta$ of the thick line in Fig. 3(c) is equal to 3.6°, almost the same as that obtained in Fig. 3(b). Since there are only 7 layers of metal and dielectric plates in total, treating the problem based on the effective medium theory can only be a rough approximation. Thus, it is reasonable to observe that the intensity of the side-lobes for the equivalent structure displayed by the thick line in Fig. 3(c) is a bit higher than that for the real structure as shown by the thick line in Fig. 3(b). Besides, the absolute intensity of the main lobe in far-field emission profile for the equivalent structure is a bit lower than that for the real structure, though not shown here. These are all generated due to the approximation treatment.

**Physical Mechanism.** Similar to the bull's eye structure, the physical principle of the beam collimation, i.e., the expansion of the beam width as observed in Fig. 2, is related to the interaction between the SP and the grating [4, 8, 15]. But the dispersion equation of the SP mode in the proposed structure should be modified as the traditional metal/air interface has regulated into an AMM/air interface. The following is a detailed understanding of the principle of our design. As it is well-known, the shallow periodical metallic grating has dual-functions similar to microwave antennas: one is to act as an inlet for transferring the incident light into the propagating SP wave and the other is to serve as an outlet for the SP wave to radiate toward the output space. Both processes follow the same phase matching condition:

$$k_0 \cdot \sin\theta \pm l \cdot 2\pi/p = k_{sp}, \quad (l = 1, 2, 3, ...) \tag{1}$$

where $2\pi/p$ is the reciprocal vector of the grating, $k_{sp}$ is the wave vector of the SP wave excited by the grooves, and $k_0$ and $\theta$ are the wavevector and emission angle of the output light (considering the outlet case), respectively. Here, it is assumed that the direction of incident light comes from or radiates toward the air space. The on-axis beam collimation with $\theta = 0°$ is a typical example of the outlet case. At such a situation, it requires that the momentum of the SP wave matches with that produced by the lattice scattering in the exit plane. With respect to the bull's eye, geometrically there is a disturbance in its exit plane, *i.e.*, a through-aperture among a set of engraved grooves, thus it requires that the excited SPs from the aperture and the grooves are approximately in-phase [8]. As a result, the distance between the slit and the nearest-grooves should be carefully designed [39]. However, in our design, as demonstrated, the nanometer-sized apertures in the bull's eye structure can be eliminated by adopting an AMM substrate. In the exit plane, the engraved grooves on the AMM substrate could excite in-phase SP wave automatically (in the absence of the slit disturbance) as displayed in Fig. 2(a). In this view, the AMM slab design allows simplified beam collimator as compared with the bull's eye.

The equity of Eq. (1) can be satisfied by adjusting the grating period. In general, it is mainly the first order of Bloch mode being excited due to the lattice scattering effect, *i.e.*, $l = 1$. For the bull's eye structure with light radiating toward the air space, the SP wave excited in the vicinity of its exit plane with engraved grooves is transformed from the eigen-mode excited on a flat metal/air interface. The dispersion relation of the surface wave on a flat metal/air interface is known as:

$$k_{sp\_flat} = k_0 \sqrt{\frac{\varepsilon_m}{\varepsilon_m + 1}}, \tag{2}$$

where $k_{sp\_flat}$ represents the wavevector of the SP wave excited on the flat interface. After the perturbation of the grooves, the wavevector of the SP wave ($k_{sp}$) is slightly larger than that on the flat interface ($k_{sp\_flat}$).

The dispersion relation of the SP wave excited on the exit plane of our structure is distinct from Eq. (2), because the shallow metallic grating is placed on top of an AMM substrate instead of an opaque metal plate. Based on the Maxwell equation and boundary conditions, we derive that the dispersion relation of the surface wave excited on a flat AMM/air interface can be expressed as:

$$k_{sp\_AMM} = k_0 \sqrt{\frac{\varepsilon_\perp (\varepsilon_\parallel - 1)}{\varepsilon_\perp \varepsilon_\parallel - 1}}, \qquad (3)$$

where $k_{sp\_AMM}$ represents the wavevector of the SP wave excited on the flat AMM/air interface. By calculation, in this case, $k_{sp\_AMM}$ is a bit greater than $k_{sp\_flat}$, thus the optimized grating period in our proposal should be a bit smaller than those adopted in bull's eye structure. For example, the grating periodicities optimized at normal emission for a typical Ag-made bull's eye structure are $0.91\lambda_0$ [7] and $0.94\lambda_0$ [22] when the incident wavelength $\lambda_0$ is 660 nm, very close to the band studies in this work. However, in our case the optimized grating period is only $0.88\lambda_0$, smaller than those in bull's eye designs, which arises from the increased wavevector of the SP wave due to the usage of the AMM substrate.

**Parameter study.** Next, we carry out the study on the performance of the proposed beam collimator when the waist radius of the incident Gaussian beam changes. Fig. 4(a) shows the normalized angular dependent far-field intensities at $w_b$ = 1.5, 0.5, and 5 $\mu$m. The results indicate that the collimation effect maintains for Gaussian beams with large waist radius. For instance, when $w_b$ = 5 $\mu$m, the half width of emission angle is the smallest, $\Delta\theta$ = 1.8 °, and the corresponding width of the output beam is the widest; besides, no side-lobes appear in its far-field profile. For the plane wave incidence, it can be regarded as a Gaussian beam with infinitely large waist radius, and its far-field emission profile produced by sufficiently enough grooves (not shown) is similar to that of $w_b$ = 5 $\mu$m. The results at $w_b$ = 1.5 $\mu$m indicates that the beam collimation process almost fails with very strong side-lobes, mainly because the grooves participate in exciting the in-phase SP wave is too few.

Finally, the effect of the geometrical parameters of the proposed structure (not the equivalent one) on the performance of beam collimation was analyzed. The calculated far-field emission profiles as functions of $p$, $h$, and the duty cycle ($w/p$) of the grating are shown in Figs. 4(b)-4(d), and the profiles as functions of the AMM slab thickness ($H_{AMM}$) and the thickness ratio of $MoO_3$/Ag ($t_2/t_1$) are plotted in Figs. 4(e) and 4(f), respectively. Among all parameters, we find that the grating period plays a dominant role in determining the beaming performance. This is reasonable because all collimation processes obey the phase matching condition in which $p$ is the determinative factor. Fig. 4(b) reveals that when $p$ is 500 nm, $\Delta\theta$ is ~7.1°, even greater than that of the incident Gaussian beam, while when $p$ increases to 600 nm, the far-field emission profile splits into two branches because the phase matching condition gets fulfilled at off-normal directions ($\theta \neq 0°$). By scanning the period with an interval of 2 nm, we find that when the structure with $p$ of 554 nm has the smallest $\Delta\theta$. When the incident wavelength is tuned, the grating period should be regulated to match the phase matching condition as the material parameters are different. Fig. 4(c) also indicates that the device performance is the best when the duty cycle ($w/p$) is 0.2. Increase in the duty cycle ($w/p = 0.3$) results in a decrease in the intensity of side-lobes, and also brings the increase of $\Delta\theta$ as shown by the star-line in Fig. 4(c), while decreasing the duty cycle ($w/p = 0.1$) not only deteriorates the side-lobes but also changes the main-lobe into a three-petal shape as shown by the triangular-line in Fig. 4(c). The far-field emission profile at $h = 30$ nm as observed in Fig. 4(d) is similar to that when $w/p = 0.1$. When the grating height is as large as 40 nm, the normalized far-field profile as shown by the star-line in Fig. 4(d) is almost the same as that when $h = 36$ nm, but calculation reflects that the transmittance is reduced to ~60% of that at $h = 36$ nm as the higher grating blocks more light. In addition, a shallow grating is desired to avoid the excitation of localized SP modes. Therefore, a suitable grating height should be chose in practical design. The width of emission angle maintains very well when the thickness of the AMM slab is tuned as shown in Fig. 4(e). Because it is difficult to fabricate a uniform continuous Ag thin film with a layer thickness thinner than 10 nm and a thick AMM slab blocks light, a 70 nm thick AMM slab was chosen. The thickness ratio of $MoO_3$/Ag affects the two effective material parameters and thus impacts on the excitation of the SP wave according to Eq. (3), however its impact on the beam collimation effect is insignificant, as indicated in Fig. 4(f).

**Conclusions**

In summary, an AMM slab-based plasmonic beam collimator, made with a stack of alternating sliver and dielectric layers and a top shallow sliver grating, has been proposed. AMM slab beam collimator has an obvious processing advantageous compared to that of the traditional bull's eye structure, removing the complex and expensive nano-fabrication steps. Considering the advantage of easy fabrication, the proposed design is a good alternative of the bull's eye in beaming application. The principle of the AMM slab beam collimation has been studied using the effective medium theory. The effects of the geometrical parameters of the AMM slab structure on the performance of plasmonic beam collimation have been simulated and analyzed systematically. As the geometric image of the proposed structure appears to be periodic along only one direction, the collimation is realized along a single axis. It is anticipated that a symmetric polarization-insensitive beam collimator can be realized by using ring-shaped grooves. The concept of the AMM based beam collimator proposed in this work could open up a viable and cost-effective approach towards beam manipulation.

**Methods**

**Simulation methods.** Light propagating behavior as well as the far-field distribution of the AMM collimator in 2D space was simulated using the Comsol software of the Finite Element Method [40]. In simulation, the grid sizes are smaller enough for generating accurate electromagnetic results. An important note for using this software is that the sign of the imaginary part of the material permittivity should be set to negative; otherwise, the simulated results would be incorrect as those reported in Ref. [41]. For simplicity, the glass substrate is considered as a semi-infinite medium in the simulation.

**Acknowledgement**

This research work was financially supported by National Natural Science Foundation of China (11204205,

61475109, 61274056, 61275037, 11204202, and 91233208). Cui also acknowledges the funding support including Outstanding Young Scholars of Shanxi Province, Hong Kong Scholar Program (XJ2013002), China Postdoctoral Science Foundation (2014M550152), Natural Science Foundation of Shanxi Province (2012011020-4), and Doctoral Program of Higher Education Research Fund (20121402120017).

**Author contributions statements**

S. Zhang and F. Zhang simulated the problems and prepared the figures with the help discussion with G. Li. Y. Cui was responsible for the original research concept and physical interpretation, and wrote the main manuscript text with S. Zhang. S. He, Y. Hao and F. Zhu helped on understanding partial of the physics and polished the paper. All authors reviewed the manuscript.

**Figure Captions**

Fig. 1. (a) Two-dimensional configuration of the proposed plasmonic beam collimator comprising of a plasmonic grating on top of an anisotropic metamaterial layer. (b) The corresponding three-dimensional view of the structure.

Fig. 2. Distributions of field $H_z$ when the Gaussian beam with 1.5 $\mu$m waist radius propagates through the proposed beaming system with alternating metal/dielectric layers (a) and through the system with bulk AMM medium (named as the "Equivalent structure") (b). The location of the beaming structure is indicated by the dotted line.

Fig. 3 (a) Angular dependent far field intensities [$I_N(\theta)$] extracted from Fig. 1D in Ref. [7] produced by the bull's eye structure (as shown in the inset, named as "Reference structure"). (b) $I_N(\theta)$ without (thin line) and with (thick line) the proposed beaming system. (c) $I_N(\theta)$ for the Equivalent structure. The half width of emission angle, $\Delta\theta$, is defined as the angular range in which the field intensity reduces from maximum to its half, as indicated in (a).

Fig. 4. Angular dependent far field intensities [$I_N(\theta)$] when the geometrical parameters of the proposed plasmonic beaming system are tuned. Unless otherwise indicated in the subplots, the default structure has $w_b$ = 1.5 $\mu$m, $p$ = 554 nm, $w/p$ = 0.2, $h$ = 36 nm, $H_{AMM}$ = 70 nm, and $t_2/t_1$ = 0.9.

**Figures**

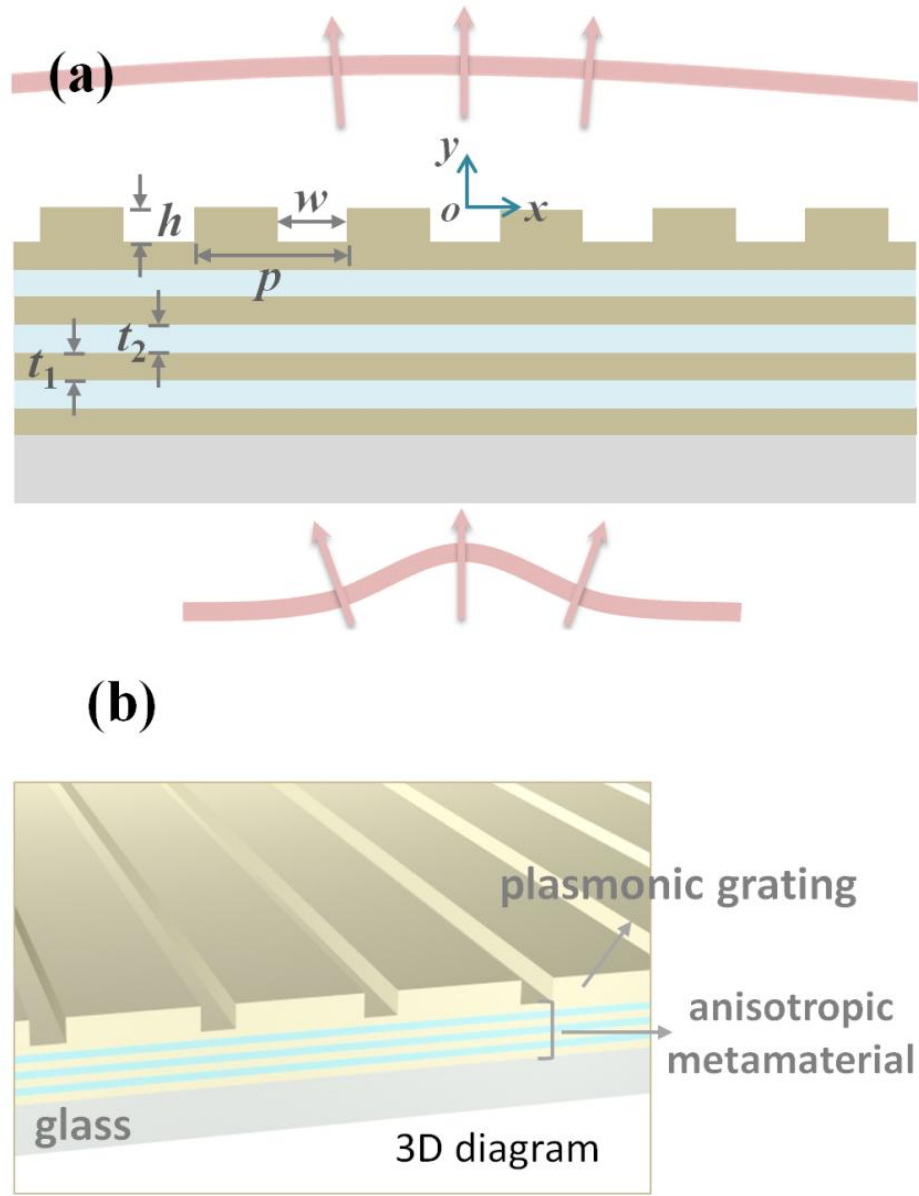

Fig. 1. Zhang, Cui, *et al.*

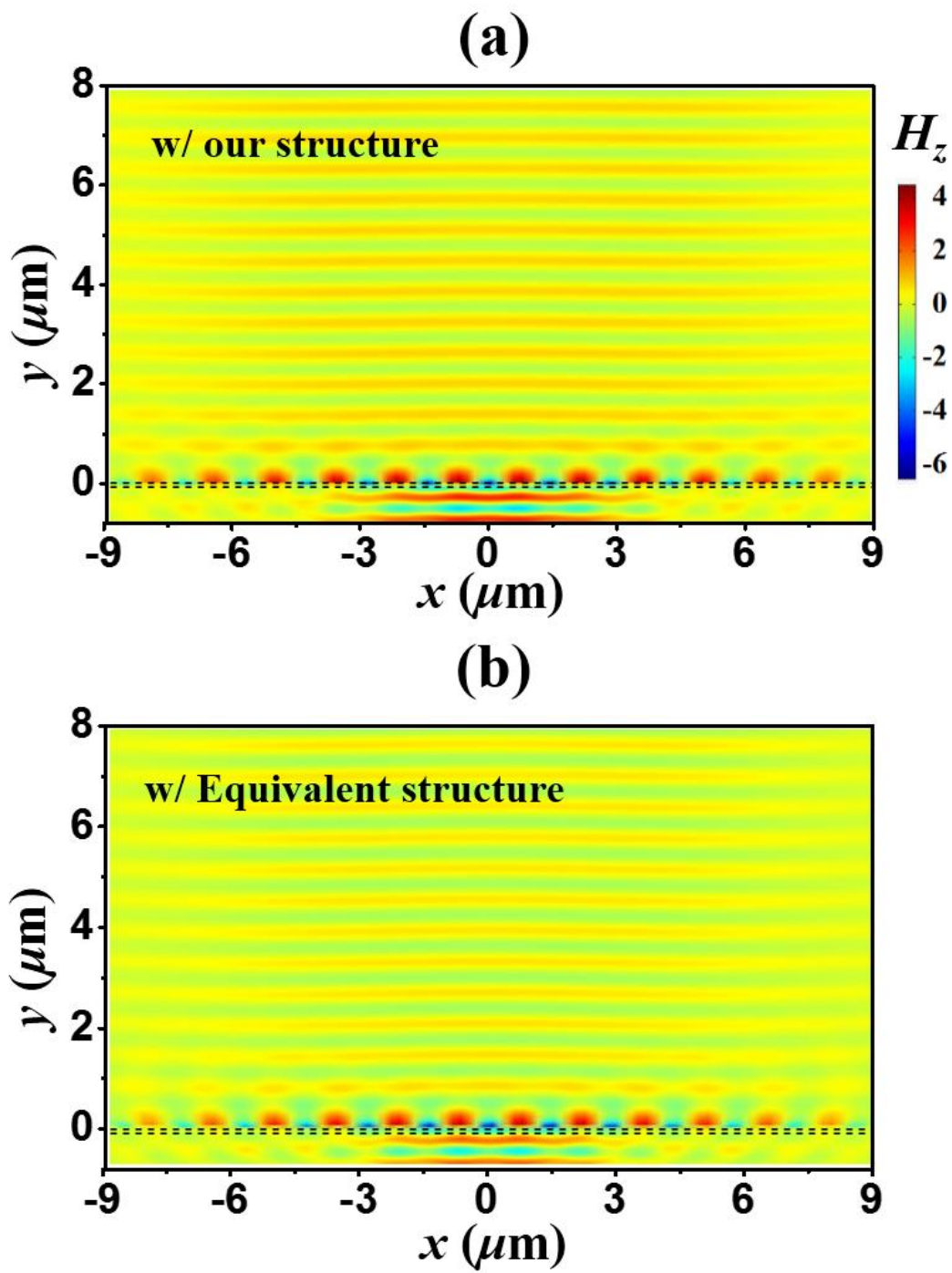

Fig. 2. Zhang, Cui, *et al.*

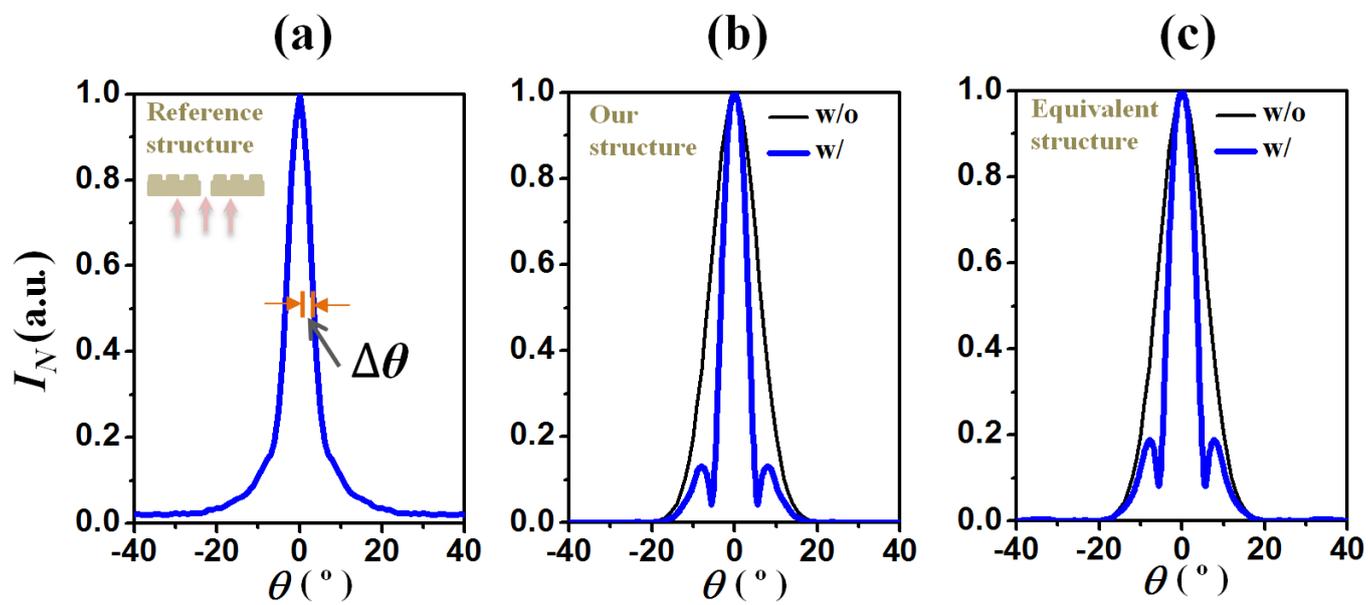

Fig. 3. Zhang, Cui, *et al.*

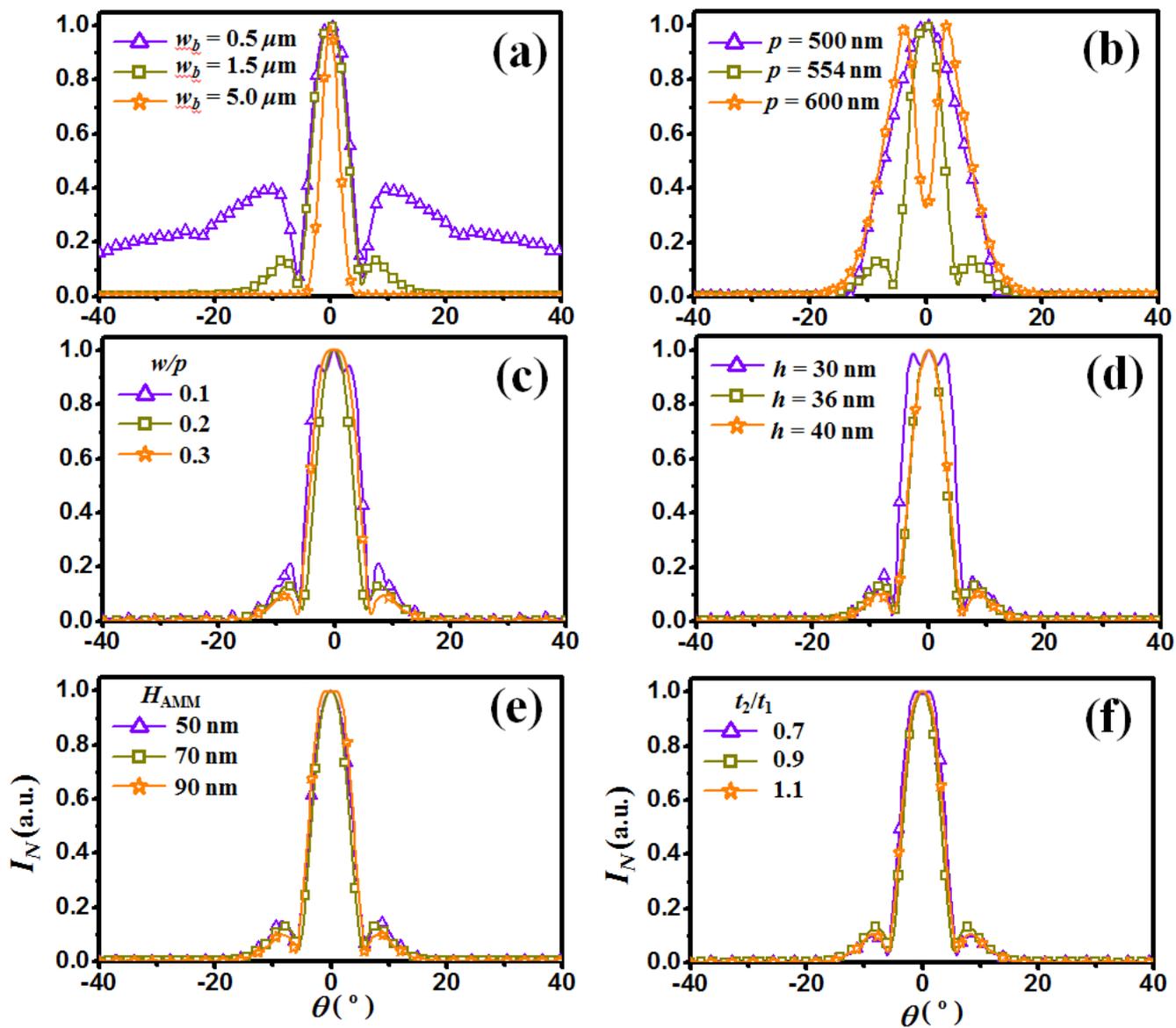

Fig. 4. Zhang, Cui, *et al.*